\documentclass[dvips,11pt]{article}
\usepackage{varioref,exscale,latexsym,amsmath,amssymb}
\usepackage{graphicx}
\def\beq{\begin{equation}}

\def\eeq{\end{equation}}

\def\beqa{\begin{eqnarray}}

\def\eeqa{\end{eqnarray}}

\title{{\bf Constraints on the variability of quark masses from nuclear binding\\}}
\author{Thibault Damour$^{a}$ and John F. Donoghue$^{a,b}$, \\ \\
$^a$Institut des Hautes \'{E}tudes Scientifiques \\
Bures sur Yvette, F-91440, France\\
and \\
$^b$Department of Physics\\
University of Massachusetts\\
Amherst, MA  01003, USA
 \\}
\begin{document}
\begin{titlepage}
\maketitle
\begin{abstract}
Based on recent work on nuclear binding, we update and extend the
anthropic constraints on the light quark masses,
with results that are more tightly constrained
than previously obtained. We find that heavy nuclei would fall apart (because
the attractive nuclear central potential becomes too weak) if
the sum of the light quark masses $m_u+m_d$ would exceed their physical values
by 64\% (at 95\% confidence level). We summarize the anthropic constraints that follow
from requiring the existence both of heavy atoms and of hydrogen.
With the additional assumption that
the quark Yukawa couplings do not vary, these constraints provide a
remarkably tight anthropic window for the Higgs vacuum
expectation value: $0.39 < v/v_{\rm physical} < 1.64$.
\end{abstract}
\vspace{0.2 in}
\end{titlepage}

\section{Introduction}

To a first approximation, the fundamental parameters that describe
our world appear to be uniform in space and constant in time.
However, it is possible that their apparent constancy is illusory
because of our limited ability to make observations across space and
time. There are mechanisms, such as chaotic inflation and the string
landscape, that can lead to a multiverse in which regions far
outside of our visible horizon have different parameters from those
that we see. Similarly it is possible (if there exists some nearly massless
``moduli field'') that the parameters could have
been different in the early universe, and there are even
experimental hints for this option. While these possibilities may or
may not bear fruit in future studies, it is important to explore
these options as carefully as possible.

The possibility of variable parameters changes the way that we
approach the open questions of fundamental
physics \cite{weinbergmultiverse,jfdmultiverse,agrawal,ArkaniHamed:2004fb}.
For example, the
existence of a multiverse with different parameters in different
domains would modify the way that we approach the issue of using
those parameters as a test of the underlying theory. Rather than
looking for a unique set of parameters to emerge from a fundamental
theory, we would expect them to be distributed in some typical range.
However, for some parameters there is a further restriction in that
there are combinations of parameters that would lead to a domain that
could not support life. While there is some fuzziness in the
constraints for the existence of life, certain clear physical
properties can be used to delineate the extreme limits of the
possible ranges. For example, {\it atoms must exist} and this restricts
the ranges of the quark and lepton masses and possibly the Higgs
vacuum expectation value (vev) \cite{agrawal}.
 This ``atomic constraint'' is particularly
significant for the Higgs vev because the small value of this
parameter (on the GUT, or Planck, scale) is one of the great
fine-tuning problems of the Standard Model and consequently it is one
of the greatest motivations for new physics.
Alternatively, taking for granted this fine-tuning changes the way one
can approach the need for new physics (and notably
supersymmetry \cite{ArkaniHamed:2004fb}).

The work of Agrawal et al (ABDS) \cite{agrawal} has used this
{\it atomic principle}  (as it was called in \cite{ArkaniHamed:2004fb}),
 i.e. the need for
the existence of atoms, to provide secure anthropic constraints
 on quark masses and the Higgs vev.  In order to translate from the
direct bound on the quark masses, this work assumes that the other
parameters of the Standard Model remain fixed while the
Higgs vev is allowed to vary. In a
realistic theory, if the Higgs vev is able to take on different
values, then the other parameters may also vary. However, the
expected range of the Higgs vev is far larger in the absence of
other new physics - this is why this vev is viewed as a great
fine-tuning problem. As a consequence this key anthropic constraint may still
have a robust meaning, even if other parameters are allowed to vary.
Within the Standard Model the quark masses follow from the weak interaction, and
are proportional to that scale. The Higgs vev sets the scale of the weak interactions.
In contrast, the major contributions
to nuclear masses are determined by the strong interactions. The general
constraint then is that
the effects of the scale of the weak interaction must overlap the scale of the strong interactions. It
is the interplay of these two very different interactions that
allows the existence of atoms. There is then a narrow volume of
parameter space that produces nuclei and atoms.

 A temporal variation of parameters could have yet different
implications. A continuous variation of some quantity implies that
this quantity is a field, i.e it carries a space-time dependence.
For this variation to occur over cosmological time scales
 the field must be nearly massless. This then suggests that such a
field coupled to matter would lead to violations of the equivalence
principle, for example, or to other observable consequences.

In this paper we use recent work on nuclear binding to address some
of these issues. In particular, we refine the understanding of the
viable range of quark masses which follows from the existence of
heavy nuclei. Other atomic constraints (namely that hydrogen exists) 
bound the possible masses of the electron.
We briefly discuss
the constraint on the Higgs vev if the
Yukawa couplings are held fixed.  We reserve to a companion paper the issue of
implications for tests of the
equivalence principle \cite{dd2}.

Our paper is organized as follows: In Section 2 we use recent work
on effective field theory to estimate which variation in quark masses
would ``unbind'' heavy nuclei. In Section 3 we provide more physical insight
into the sensitivity of nuclear binding to scalar interactions between nucleons
by considering a simple model for homogeneous nuclear matter. Finally, Section 4
displays the anthropically allowed range of the masses of the first generation of
quarks and leptons: $m_u, m_d, m_e$.

\section{Anthropic constraints from the existence of nuclei using effective field theory}

To the extent that we understand how the Standard Model leads to the
physical world that we observe, we should  be able to
understand how that world would change if we modify the parameters
of the theory to take on values in the neighborhood of their
physical values. While we feel that we do understand the overall
phenomenology of the Standard Model, the precision that we claim in
these calculations continues to advance at a relatively modest pace,
especially for what concerns the link between the fundamental Lagrangian
and nuclear physics.
However, the recent advances in nuclear physics have been
impressive, largely through the application of the {\it effective field
theory approach} \cite{eft}. Since the energies in nuclear processes are low, the
effective field theory framework parameterizes the key ingredients in
terms of a relatively small number of low energy constants. This method has
been applied extensively to nuclear binding and has put traditional nuclear
phenomenology on a more solid basis.

We consider a relatively simple but reasonably model-independent description of
the parameters that influence nuclear binding, limiting ourselves to
those that appear most important. For all but the lightest of
nuclei, the key aspect of binding comes from a central potential
that is isospin symmetric and which does not involve the spin of the
nucleons. These will be parameterized by a small number of contact
interactions \cite{furnstahl, serot, walecka}. While the other components of the nuclear force are
important for the detailed descriptions of nuclear states, the main
contributions to the binding energy comes from this spin-singlet  and
isospin-singlet central potential.

In addition to this model-independent framework, we employ the results of recent work
on the variation of the dominant coupling constants with a
changing quark mass \cite{chiral}.
While there are clearly some uncertainties in this calculation, it is easy
to argue that the dominant effects are kinematic. The coupling constants are
calculated using a dispersive representation \cite{vinhmau, sigma}, with the threshold of the dispersion
integral appearing at the physical threshold of $2 \, m_\pi$. Raising the threshold is
seen to lead to a kinematic suppression of the coupling strength. While our estimate
is much more sophisticated than this, nevertheless the dominant effect is that of the
kinematic threshold.

In effective field theory, the propagation of the very light degrees
of freedom must be treated dynamically because these particle can
propagate long distances. By contrast, at low energies, the more massive
degrees of freedom cannot propagate far and can be represented by
contact interactions - i.e. delta function potentials and
derivatives of delta functions. This has the effect of simplifying
the contributions of various possible particle exchanges, with
various spatial potentials, into a few low energy constants
describing the strength of the interactions. In nuclear processes,
it is useful to treat the direct effects of one-pion exchange
dynamically, but to treat the other components of the nuclear force
by contact interactions. For the spin singlet and isospin singlet
central potential responsible for nuclear binding there are then two
possible contact interactions, called  {\it scalar} and {\it vector}
\begin{equation}
H_{\rm contact} = G_S (\bar{N}N)(\bar{N}N) + G_V (\bar{N}\gamma_\mu
N) (\bar{N}\gamma^\mu N) +...
\end{equation}
where $N$ denotes the nucleon field, and
where $G_S$ is negative (i.e. attractive), while $G_V$ is positive (i.e. repulsive).
In traditional meson exchange models, the scalar component
corresponds to the exchange of the $\sigma(600)$ meson and the
vector component corresponds to the exchange of the $\omega(783)$
meson.

Our first task is to understand the primary ingredients of nuclear
binding in this framework. Fortunately the dominant ingredients
in the binding of heavy nuclei have been
elucidated in a set of papers by Furnstahl, Serot and co-workers
\cite{ furnstahl, serot,walecka}. For heavy nuclei, one-pion-exchange is not very
important because pion exchange is proportional to the spin and
isospin operators and the spins and isospins of most nucleons
average to a total that is close to zero. Instead the isoscalar and
spin independent contributions sum over all nucleons and are
dominant once one is away from the few-nucleon cases. This is in accord
with the standard wisdom that the nuclear central interaction ($J=0$ and $I=0$)
is responsible for nuclear binding. The results
for heavy nuclei can be extracted from Fig. 1 and Fig. 2 of
\cite{furnstahl}. As expected, the dominant effects are the
scalar and vector contributions described above. Other interactions
play reduced roles, although for a complete understanding of the
binding about a half-dozen contact interactions are required. Here we
will  focus our attention on the dominant isoscalar-scalar and
isoscalar-vector interactions.

Using Ref.~\cite{furnstahl}, one can quantify these
contributions to nuclear binding. We parameterize the results in
terms of the strengths of the contact interactions, normalized to
their physical values, defining
 \begin{eqnarray}
\eta_S &\equiv& \frac{G_S}{G_S|_{\rm physical}}
\nonumber \\
\eta_V &\equiv& \frac{G_V}{G_V|_{\rm physical}}
\end{eqnarray}
The contributions to the binding energy (B.E) for $^{16}O$ (in
MeV)\footnote{Though we shall use here for convenience the usual
physical units MeV (or  GeV), one should think of these (when considering
variations of the quark masses) as being defined
as some pure number times the chiral limit of the QCD confinement scale, say
$\Lambda_{\rm QCD}^{(0)}$.} are
\begin{equation}
\label{O}
{{\rm B.E.}\over A} \simeq -82 \eta_S + 44\eta_V + 30
\end{equation}
where $A$ denotes the total baryon number.
The first two terms are the effects of the scalar and vector
isoscalar interactions. The third term is the sum of all other smaller
contributions to the binding energy and kinetic energy
contributions. There is in addition the Coulomb energy and a small
center of mass correction. For $^{208}Pb$, the result is
\begin{equation}
\label{Pb}
{{\rm B.E.}\over A} \simeq -104 \eta_S + 57\eta_V + 36
\end{equation}

The results of these calculations can be generalized to other nuclei
by a parameterization that resembles the semi-empirical mass
formula. For local interactions, because the nuclear density is
nearly constant in the central region one expects that the binding
energy will have a dependence on the volume, which in turn is proportional
to the number of particles, $r^3\sim A$, and that interactions that
occur near the nuclear surface would have a modified result
proportional to the number of nucleons near the surface, $r^2\sim
A^{2/3}$. This suggests that binding effects can be parameterized in
terms of behavior in $A$ and in $A^{2/3}$. Using the results for
nuclear matter and for specific nuclei, we find a good fit of the
form
\begin{equation}
{{\rm B.E.}\over A} = -(120 -\frac{97}{A^{1/3}}) \eta_S + (67 -
\frac{57}{A^{1/3}} )\eta_V  +{\rm residual ~terms}
\end{equation}

The primary difficulty in applying these ingredients to anthropic
constraints is the need to connect the contact interactions to the
fundamental parameters of QCD. However, there is two decades worth
of work exploring the ingredients in this connection. The framework
used below follows \cite{chiral, sigma} in employing
dispersion relations, which can be used to express
the desired couplings as integrals over reactions involving
physical intermediate states. The low energy portions of these reactions
can be well predicted by chiral perturbation theory, in which one
has reasonable control over the quark mass dependence.

In general, an effective field theory prediction would be expected to
have the following structure. The high energy end of a dispersion integral
would be expected to depend on the quark masses only weakly. This is known from
the dependence of hadron masses and couplings on the quark mass parameters.
For example, if the $u,~d$ masses were doubled (keeping $\Lambda_{\rm QCD}^{(0)}$
fixed) the nucleon mass would increase by about
a half a percent. However, the low energy portions of a dispersion
integral can have a much greater change. For example, the doubling of the $u,~d$ masses
would raise the energy threshold in the dispersion relation by 40\%
and would forbid any contributions below this new threshold. In this case,
a reasonable first approximation to an effective field theory calculation
would be to treat the high energy portion of the dispersive integral
as being independent of the masses and to calculate the low energy
effects using chiral perturbation theory. Any large dependence
on the light quark masses should come from the low energy end.
This is the result of our work.

The reasoning above suggests that the most important effect is in
the scalar channel. This is the only portion of the central force that
receives large effects from low energy, as two pion exchange is
the most important contribution\footnote{The vector channel has also been
explored in \cite{chiral} but has little low energy effect and
only a very small mass dependence. We will include it in our numerics below,
while focussing, in the text, on the dominant scalar-channel effects.}.
This channel has been explored in great
depth within the context of chiral perturbation theory, including
studies very similar to the approach used in this paper \cite{egm}. One of the
authors has recently extended this work to include the constraints
of unitarity \cite{chiral, sigma}. The result is a description of two-pion exchange that
carries the main properties needed for the scalar central
potential.
We will employ this work in our analysis below. In this work, we
use chiral perturbation theory at low energies and also attempt to
extend the description to high energy. The low energy part is
then model-independent while the high energy portion is less rigorous.
However the high energy portion plays little role in our answer, since
it conforms with the expectation that it should be largely independent of the
quark masses. Moreover, we should
note here that the primary ingredient is independent of the details
of this calculation. The general trend is inescapable -
as the pion mass gets larger, the effect of two pion exchange must
get smaller. In the chiral framework, the connection of the quark
masses to the two pion threshold is well defined, and as noted above,
most of the
effect found in Ref. \cite{chiral} is kinematic.

Let us summarize the results of \cite{chiral} and extend them to
larger values of the pion mass. First, it was found that the pion mass
dependence of omega exchange (corresponding to the {\it vector} channel)
is of ``normal'' size, i.e. $O(m_\pi^2/(1{\rm GeV})^2)$. Such a ``normal'' sensitivity
to $m_\pi^2$ (and therefore to quark masses) leads to sub-leading corrections
compared to the effects linked to the  $m_\pi^2$ sensitivity of the
scalar channel. Indeed, because of the dependence on the two pion
threshold, the scalar contact interaction is much more sensitive to the
pion mass. In full generality, one has the sum rule
\begin{equation}
G_{S} = \frac{2}{\pi}\int_{2 \, m_\pi}^{\infty} \frac{d\mu}{\mu}
~{\rho}_{S}(\mu)
\end{equation}
where $\rho_S(\mu)$ is the spectral function that describes the
physical two pion intermediate state at energy $\mu$. The dependence
of this spectral function on the quark masses is explored in detail in Ref.
\cite{chiral}. The rise of the amplitudes from the threshold value
of $\mu = 2 \, m_\pi = 270$~MeV is tempered at higher energy by
unitarity effects such that the main contributions come from
energies near $\mu=500-600$~MeV. When changing the quark masses, all
ingredients change to some extent. However the key effect is the
threshold behavior. In lowest order, the pion mass-squared is
proportional to the light up and down quark masses
\begin{equation}
\label{mpimq}
m_\pi^2 =  B_0 (m_u + m_d)
\end{equation}
where $B_0$ is a constant\footnote{The precise value of $B_0$ depends on the
renormalization scale used to specify the quark masses.} (proportional to $\Lambda_{\rm QCD}^{(0)}$).
 The evidence is that this relation holds
throughout the region of interest to us here \cite{Leutwyler}.
The higher threshold then cuts off the effect of two pion exchange as
the pion mass increases.
\begin{figure}[ht]
 \begin{center}

\includegraphics[scale=1.0]{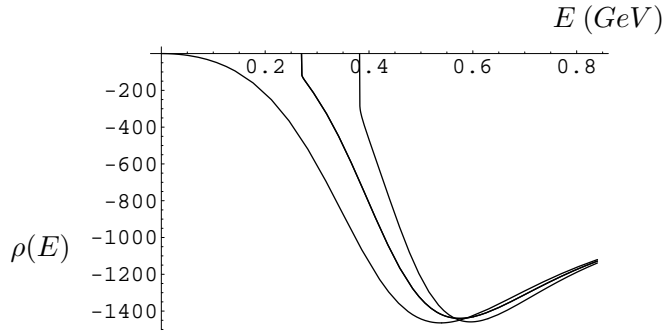}
 \end{center}
 \caption{\small{The scalar spectral function for three values of the pion mass,
 $m_\pi = 0, m_{\rm phys}$, and $\sqrt{2} m_{\rm phys}$, with thresholds starting at
 $\mu=2 \, m_\pi$ . }}
 \label{spectral}
\end{figure}

In detail, the framework of Ref. \cite{chiral} includes all variations
in the parameters governing two-pion exchange, including $g_A$, $F_\pi$ and the
$\pi\pi$ rescattering amplitude. While that work was focussed on the situations
where the pion mass was lighter than its physical value, the framework also extends
to larger values of the pion mass. For example, the comparison of the result at
the physical mass to the case where the pion mass is 40\% larger than the physical value
is shown in Fig. 1. The spectral integral will clearly show a {\it decrease} when the
pion mass is {\it increased}.

\begin{figure}[ht]
 \begin{center}
  \includegraphics[scale=1.0]{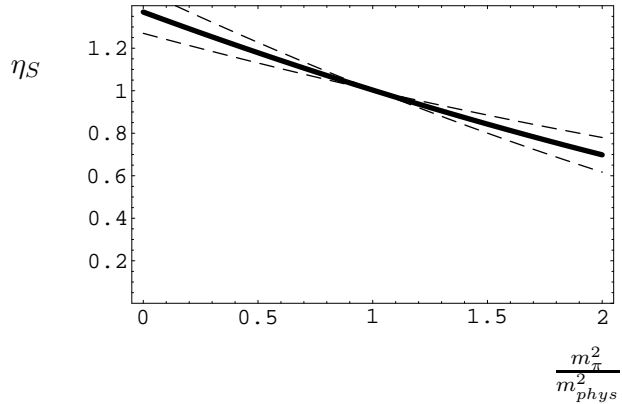}
 \end{center}
 \caption{\small{The value of the scalar strength $\eta_S$ as a function of the pion mass. }}
 \label{spectral}
\end{figure}

In \cite{chiral} it was found that the scalar strength $G_S$ reached,
 in the chiral limit, the larger\footnote{In absolute value; let us indeed recall
 that $G_S$ is {\it negative}.} value
\begin{equation}
\frac{G_S|_{\rm chiral}}{G_S|_{\rm physical}} = 1.37 \pm 0.10
\end{equation}
The error bar comes from the limitation of our understanding of the
dependence of various couplings on the pion mass.
This result could be used by itself to reasonably extrapolate to larger values of the
pion mass since the extrapolation is almost linear in $m_\pi^2$. However, there are some
non-linear features. In practice, a more detailed calculation, including required non-analytic
contributions yields the result shown in Fig. 2 for $\eta_S$, i.e. the value of $G_S$ normalized to the physical
value, as a function of the pion mass. The estimates of
these uncertainties are also shown in Fig. 2. The error bars come from from our lack of understanding
of the dependence of some of the pion and nucleon parameters on the value of $m_\pi$. The largest source
of uncertainty is the mass dependence of the axial coupling $g_A$. These uncertainties are discussed
in more detail in \cite{chiral}.

In this calculation we have calculated the spectral integral up to an
energy of 850 MeV. This includes some energies above the scale where the
chiral perturbation theory description is valid - the upper end of this integral
is modeled by using the continuation of the unitarized chiral amplitudes above the region
where they are known to be correct. However,
because very little mass variation is seen in the upper energy region, there
is an alternate procedure which does not make this model-dependent assumption yet
which yields essentially the same result. In this procedure, one calculates the
spectral integral only in the region where the chiral expansion is valid, for
example up to an energy of $600$~MeV, and includes a short distance contact interaction
to account for the effects of higher energy. (This rationale is described in more
detail in \cite{chiral}.) If one assumes that the mass dependence of the
short distance effect is of normal size (i.e. of order $m_\pi^2/ (1 {\rm GeV})^2$), then
essentially all the mass variation comes from the low energy end, reproducing the result
quoted above within error bars.

Because the effect of the scalar interaction is attractive ($G_S <0$) while the
effect of the vector interaction is repulsive ($G_V >0$), there is a substantial
cancelation between these two effects (see next Section for an analytical discussion
exhibiting this cancelation). The mass dependence of the
vector interaction has also been estimated in \cite{chiral}. It has no significant
threshold dependence because the dominant feature - the $\omega$ meson - is a narrow
pole with only a small dependence on the quark masses. We have taken this into
account in our numerics, but do not discuss it further here.
\begin{figure}[ht]
 \begin{center}
  \includegraphics[scale=1.0]{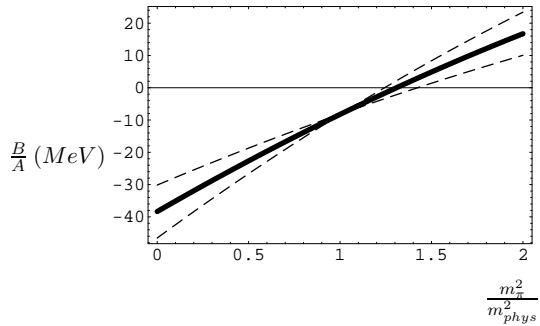}
 \end{center}
 \caption{\small{The binding energy per nucleon in $^{16}$O as a function of the pion mass. The corresponding result in $^{210}$Pb is very similar. }}
 \label{Bindingenergy}
\end{figure}

Because the scalar strength has significant variation while the vector one is less
affected, the cancelation between the two has an even larger percentage variation.
In particular,
as the {\it attractive} scalar interaction becomes weaker, it no longer dominates
over the {\it repulsive} vector interaction, and the binding energy can
change sign (from ``binding'' to ``unbinding'') as $m_\pi^2$ increases above
its physical value. Using the results (\ref{O}), (\ref{Pb})
quoted above, we see that the binding energy vanishes for a scalar strength only 10\%
smaller than the physical values
\begin{eqnarray}
\eta_S|_{\rm critical} &=&  0.90 ~~~~~~ {\rm for}~^{16}O  \nonumber \\
\eta_S|_{\rm critical} &=&  0. 89 ~~~~~~ {\rm for}~^{208}Pb
\end{eqnarray}
Study of the general formula shows that these values are typical of the whole range in
$A$. As we have seen above, increasing the pion mass will lead to a decrease in the scalar
strength. In
Fig. 3 we show the resulting nuclear binding for $^{16}O$ as a function of the
pion mass, including the estimated error bar. In producing this
figure we have assumed that the other small contributions to the
binding formula do not have significant variations. We see that this element becomes unbound
when the pion mass-squared is $36 \pm 14 \%$ larger than the physical value. This critical value
is almost independent of the value of $A$.

The anthropic constraint on quark masses can be inferred
from these results. Using the basic relation (\ref{mpimq})
between the pion mass and the quark masses, one obtains the
constraint
\begin{equation}
\frac{m_u+m_d}{(m_u+m_d)_{\rm phys}} < 1.36 \pm 0.14  ~~,
\end{equation}
from the requirement that nuclear binding exist at all.
To the best of our present understanding of pion physics from chiral
studies and from lattice simulations, the corrections to the basic
relation between pion and quark masses are negligible compared to the other uncertainties in the
calculation. If we had used the binding of
$^{208}Pb$ we would have obtained essentially the same constraint on the pion mass. The use of the semi-empirical mass
formula described above says that this constraint is roughly independent of the value of $A$.
If we include the error bar and convert to a 95\% confidence level upper bound
we conclude that
\begin{equation}
\label{finalbound}
\frac{m_u+m_d}{(m_u+m_d)_{\rm phys}} < 1.64 ~~.
\end{equation}
We will use this as our final ``atomic bound''.

\section{Constraints using a model for nuclear matter}

In this section we use a simple model for nuclear matter to provide more physical insight into the sensitivity
of nuclear binding to the scalar strength and to reinforce the results of the previous section.
The model is a variant of the description of nuclear matter discussed in
Ref. \cite{walecka} using nucleonic and mesonic fields. It reproduces the dominant contact interactions used above and
also includes higher order dependencies on the scalar couplings and the kinetic energy. We will see that these
higher order dependencies increase the sensitivity to $G_S$ and hence to the quark masses.

The starting Lagrangian is
\begin{equation}
L = \bar{\psi}\left[i\gamma^\mu\partial_\mu - g_V V_\mu \gamma^\mu -(M-g_S \phi)\right]\psi +\frac12 m_V^2 V_0^2 -\frac12 m_S^2 \phi^2
\end{equation}
where $\psi$ is the nucleon field, $\phi$ is a scalar, isoscalar field (``the sigma'') and $V_\mu$ is an
isoscalar vector field (``the omega'').

We now consider the effect of this Lagrangian in an infinite nuclear medium.
The nucleon field fills the available states up to the
Fermi energy. The density of nucleons is given by
\begin{equation}
\rho_B = \frac{\gamma}{(2\pi)^3}\int_0^{k_F} ~d^3k = \frac{\gamma k_F^3}{6\pi^2}
\end{equation}
where $\gamma$ is the number of degrees of freedom ($\gamma=4$ for isoscalar nuclear
matter, which we will use in our numerical work) and $k_F$ is the Fermi momentum.
The nucleon field acts as the source of the scalar and vector fields. Solving for the
energy density of this uniform distribution, one finds
\begin{equation}
\epsilon = \frac12 \frac{g_V^2}{m_V^2} \rho_B^2+\frac12 \frac{g_S^2}{m_S^2} \rho_S^2 + \frac{\gamma}{(2\pi)^3}\int_0^{k_F} ~d^3k ~E^*(k)
\end{equation}
where the scalar density is
\begin{equation}
\rho_S = \frac{\gamma}{(2\pi)^3}\int_0^{k_F} ~d^3k \frac{M_*}{E^*}
\end{equation}
and
\begin{eqnarray}
E^* &=& \sqrt{k^2+M_*^2}  \nonumber \\
M_* &=& M - g_s \rho_S
\end{eqnarray}
We approximate this by non-relativistic kinematics, which is a reasonable approximation for
nuclear matter. If we then
solve for the energy per baryon ($E/A$, which is $\epsilon/\rho_B$) we find
\begin{equation}
\frac{E}{A} -M =  \frac{\gamma}{12 \pi^2} k_F^3 (\frac{g_V^2}{M_V^2}-\frac{g_S^2}{M_S^2} )
+\frac{3}{10}\frac{k_F^2}{M(1-\frac{\gamma g_S^2 k_F^3}{6\pi^2 m_S^2 M})}
\end{equation}
Here in the first term we see the effects of the scalar and vector contact
interactions (with $G_S =-g_S^2/m_S^2$ and $G_V=+g_V^2/m_V^2$),
while the second term is the kinetic energy of the nucleons propagating in the nuclear medium. In the language of
Ref. \cite{furnstahl}, these latter
terms would be described as higher order contributions in the kinetic energy term.

If the couplings are chosen appropriately, one reproduces the existence of nuclear matter. As the density
($\propto k_F^3$) increases, the kinetic energy initially gives a positive contribution which is eventually overcome by the potential energy
(if $g_V^2/M_V^2-g_S^2/M_S^2= G_V + G_S$ is sufficiently negative), with nuclear saturation seen in the existence of a minimum in the potential
energy function. Using appropriate values ($G_S = - 362$~ GeV$^{-2}$ and $G_V = 270$~ GeV$^{-2}$)
an energy function very similar to that of \cite{walecka} is shown as
the bottom curve in Fig 4, reproducing the correct binding energy and Fermi momentum.

\begin{figure}[ht]
 \begin{center}
  \includegraphics[scale=1.0]{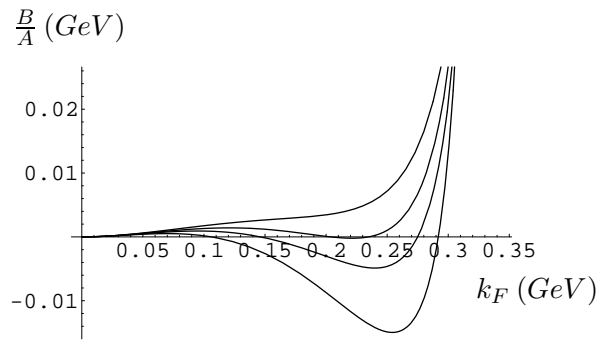}
 \end{center}
 \caption{\small{The binding energy, $B$, per nucleon in nuclear matter as a function of the Fermi momentum for various values
 of the scalar strength $G_S$. From bottom to top in the figure, the values of $- G_S$ are (362, 340, 328, 315) GeV$^{-2}$}}
 \label{BAkF}
\end{figure}
\begin{figure}[ht]
 \begin{center}
  \includegraphics[scale=1.0]{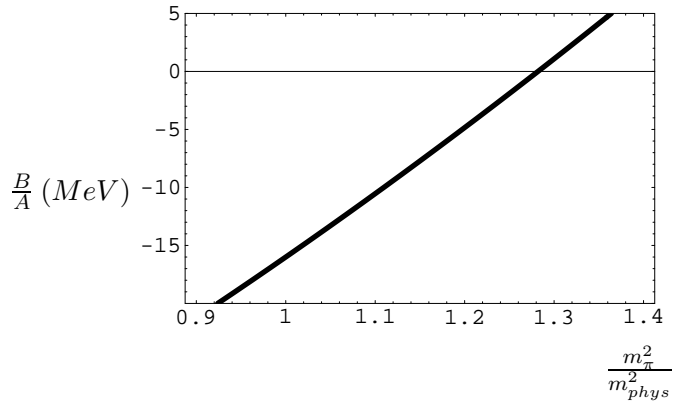}
 \end{center}
 \caption{\small{The binding energy, $B$, per nucleon in nuclear matter as a function of the pion mass.}}
 \label{BEmatter}
\end{figure}

Now let us consider variations in the scalar coupling. Various other values of $G_S$ are also shown in Fig. 4.
We observe that the binding is highly sensitive to the scalar coupling. In particular, nuclear matter disappears
for a critical value of $G_S$ only 10\% smaller than the physical value
\begin{equation}
\eta_S|_{\rm crit} = \frac{G_S|_{\rm crit}}{G_S|_{\rm phys}} = 0.904
\end{equation}
This confirms the sensitivity to this parameter found in finite nuclei by Ref. \cite{furnstahl}. In fact, we can see
from both the formula and the numerics that higher order effects $G_S$ have the effect of making the sensitivity greater.

Finally let us translate this into a constraint on the quark masses. Using the calculation of
$G_S$ as our guide, the binding energy of nuclear matter as a function of the pion mass is shown in Fig. 5.
We see that the central value of the constraint satisfies
\begin{equation}
\frac{m_u+m_d}{(m_u+m_d)_{\rm phys}} < 1.28 \pm 0.14
\end{equation}
This is completely consistent with, and slightly stronger than, the bound quoted in the previous section.
Because the two constraints overlap, to be conservative we will use the upper bound of the previous
section as our final constraint.

\section{Summary of quark and lepton mass constraints}

In this section, we display the anthropically allowed range of the masses of the first generation of quarks and leptons, $m_u,~m_d,~ m_e$,
updating Ref. \cite{hogan}. There are two primary constraints.
One is a bound on the sum of quark masses $m_u+m_d$ derived above. If this combination becomes too large, all nuclei
fall apart because the attractive central potential becomes too weak. The other bound follows from the constraint
that if the neutron mass is lighter than the sum of the masses of the proton and electron, hydrogen will be unstable through
the capture of electrons
$e^- + p \to n + \nu$, such that a hydrogen atom will decay\footnote{We assume throughout that the neutrino mass, if it is indeed allowed to also vary, remains negligibly
small. There is, moreover, an anthropic constraint that
ensures this result\cite{tegmark}.}.
In practice, these two constraints
suffice to provide tight bounds on these three
masses.

For the first constraint due to our bound following from the binding of nuclei, we need to express this in terms of absolute masses.
While our constraint, Eq. (\ref{finalbound}), involves the ratio of masses, which is scale independent, the absolute masses depend on the scale
that they are specified at. The most canonical values of the quark masses $m_d\sim 7$~MeV and $m_u\sim 4$~MeV are typically taken to
apply at a scale of 1 GeV, and we will use this prescription. In this case, the bound on the ratio, Eq.~(\ref{finalbound}), implies that
\begin{equation}
m_u +m_d \le 18 ~{\rm MeV}~~.
\end{equation}

The second constraint - that hydrogen exists\footnote{The existence of hydrogen is probably
necessary for having long-lived stars, quietly shining for eons and thereby providing a favorable
environment for the appearance of life. Hydrogen might also be necessary  to
make biological molecules.} - involves a bound on the physical masses
\begin{equation}
m_P+m_e \le m_N~~.
\end{equation}
If this relation is violated, the electron in the hydrogen atom will be captured by the proton\footnote{The violation of this relation
will also cause important modifications in heavy nuclei. However, bound protons can still exist in heavy atoms if they are sufficiently
more deeply bound than neutrons, such that the Pauli principle blocks the proton to neutron conversion. We do not attempt to
analyze this situation in detail, using instead the simpler constraint on hydrogen as the main anthropic bound.}. 
To convert this relation to the quark level we need to estimate both the quark contribution to the neutron-proton
mass difference and the electromagnetic contributions. Let us parameterize these by
\begin{equation}
m_N-m_P = Z_0(m_d-m_u) - \epsilon_{EM}
\end{equation}
Here the first term on the right hand side is the contribution due to the differences in quark masses, while the second part is the
electromagnetic contribution to the mass difference. Since the quark masses are scale dependent, so also is $Z_0$, such that
the product is scale independent. Both potential models \cite{isgur} and bag models \cite{deshpande} yield remarkably
similar values for the electromagnetic contribution, $\epsilon_{EM} \sim 0.5 $~MeV. The use of the canonical values of the masses
at a scale of 1 GeV then implies that $Z_0= 0.6$ in order to obtain the correct neutron-proton mass difference.
This is a
very reasonable value and we will adopt it in our numerics. Using these values, we find that the difference in quark masses is also bounded
\begin{equation}
m_d-m_u \ge \frac{m_e + \epsilon_{EM}}{Z_0}
\end{equation}
or
\begin{equation}
m_d-m_u - 1.67 m_e \ge 0.83 ~{\rm MeV}~~.
\end{equation}
The right hand side of this latter constraint is evaluated at the physical value of
the fine structure constant and the $QCD$ scale. It is linear in both of these
quantities.
\begin{figure}[ht]
 \begin{center}
  \includegraphics[scale=1.0]{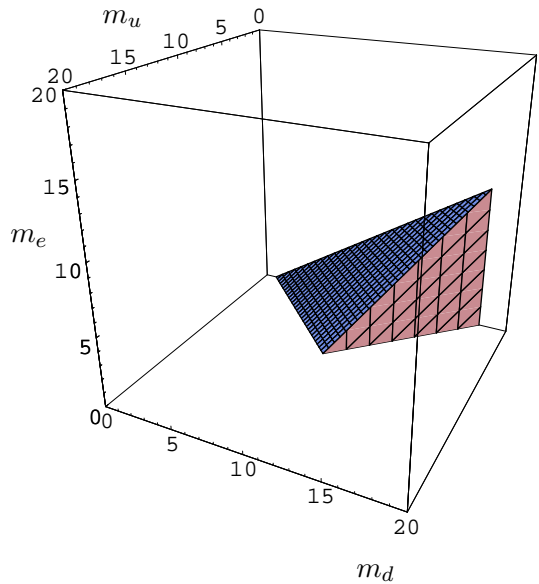}
 \end{center}
 \caption{\small{The anthropic constraints on $m_d, ~m_u,~m_e$ in MeV units.}}
 \label{3Dconstraints}
\end{figure}

\begin{figure}
\begin{center}
\begin{tabular}{ccc}
{\includegraphics[scale=0.5]{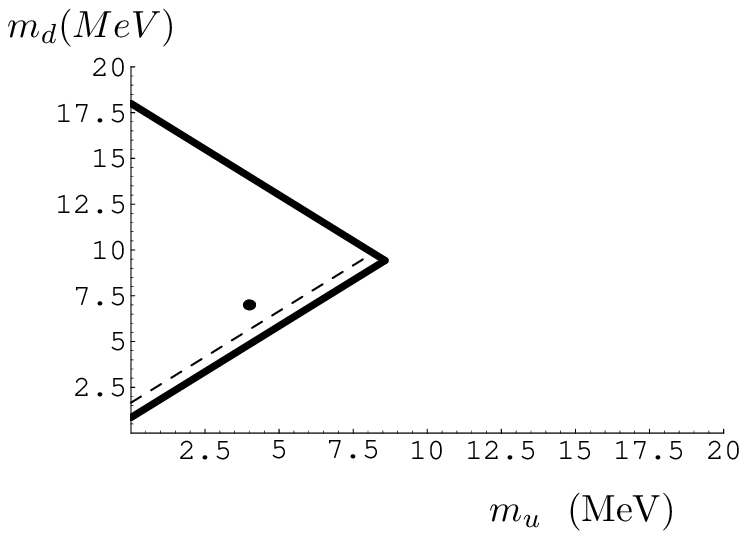}}&
{\includegraphics[scale=0.5]{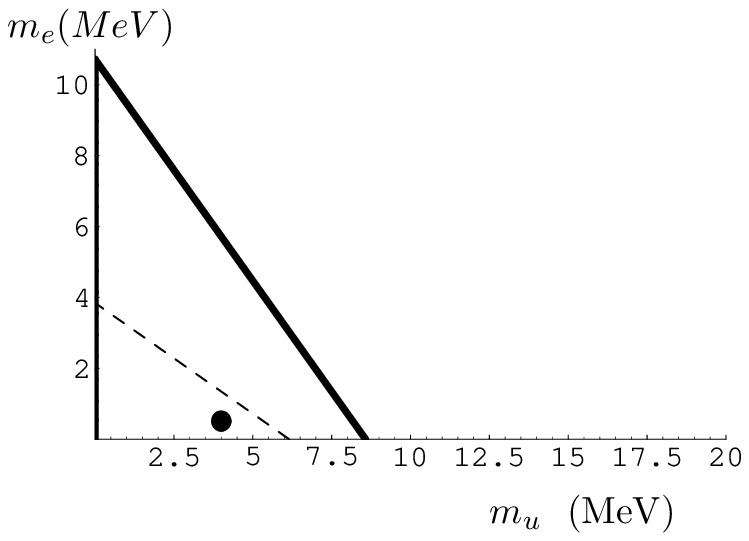}}&
{\includegraphics[scale=0.5]{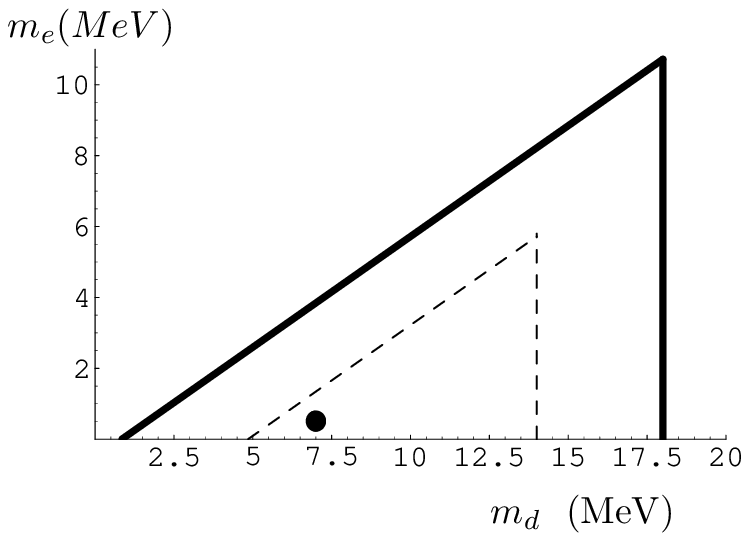}}
\end{tabular}
\end{center}
 \caption{\small{The projection of the anthropic constraints of Fig. \ref{3Dconstraints} into the planes of each pair of masses. The
 solid lines denote the total allowed region, while the dashed line shows the remaining 
 area if the third mass takes on its physical value. The dot
 shows the physical values of the masses.}}
 \label{projconstraints}
\end{figure}
The constraints (20) and (24) are plotted in Fig. 6, which shows a 3D plot listing the allowed values of each combination of mass. The important point is that the
two constraints manage to {\it provide bounds on all three of the masses}. Note that $m_u$
and $m_e$ have no lower anthropic bounds, while
$m_d$ is constrained to be non-zero.

We can also take projections into various two-dimensional subsections. The constraints on various combinations of the masses are
shown in Fig 7. In each case, the outer range is shown allowing the third mass parameter to take on any allowed value.
Also marked by a dashed line on these plots is the overall range of the masses when the third mass parameter takes on its
physical value.

We see that quite small changes in the quark masses would lead to unlivable conditions.

These ranges for the masses can be converted to an allowed range for the Higgs vacuum expectation value, under the additional assumption that
 the other parameters of the Standard Model (Yukawa and gauge couplings) are held fixed. This extra assumption could possibly
occur in grand unified theories, where there are many attempts to predict gauge and Yukawa couplings. In this situation, the Yukawa couplings
could be fixed by the symmetries of the grand unified theory, and our quark mass constraints translate directly into constraints on the
Higgs vev. From our work above on the binding of nuclei we would then find
\begin{equation}
\frac{v}{v_{\rm phys}} < 1.64
\end{equation}
at 95\% confidence.
This constraint is both stronger than and independent from the final
result of ABDS \cite{agrawal}. The latter was based on the fact that
as the quark masses increases, at fixed Yukawa couplings, the
neutron-proton mass difference increases until eventually all bound
neutrons decay and only protons exist. Thus, that bound constrains
$m_d - m_u$, while ours constrains $m_d + m_u$. Moreover, our present bound
is tight enough that it supersedes the bound on the mass difference, because
$m_d-m_u$ can never be greater than $m_d+m_u$.
The lower constraint comes from the other
process discussed in this section - the stability of hydrogen atoms against the reaction $p+e \to n +\nu$. If the Higgs vev becomes too small, the
proton becomes heavier than the neutron due to electromagnetic interactions and this reaction occurs. Since the up, down and electron masses
are all proportional to $v$, one finds that this constraint is
\begin{equation}
\frac{v}{v_{\rm phys}} \ge 0.39
\end{equation}
When combined one finds a very restricted range for the vev, under the stated assumptions:
\begin{equation}
0.39 \le \frac{v}{v_{\rm phys}} \le 1.64
\end{equation}
which is especially tight if one considers it in the the context of Grand Unification, where the natural range for the vev could extend up to the
GUT scale.

Of course, it is also possible that the extra assumption about the constancy of the Yukawa couplings is not correct. In the
discussions of the string landscape, there are so many possible vacua that others with different values of the Yukawa couplings
should be possible. However, our quark mass constraints should still be relevant for describing the likely values of the
Higgs vev \cite{inprep}. Even though extreme cases with disparate scales may be possible \cite{kribs}, it is plausible that the
need for light quarks makes it likely that the Higgs vev is close to the scale of the strong interactions \cite{inprep}.
Moreover, in theories such as supersymmetry which use dynamics to stabilize the fine tuning problem, the anthropic constraint could
be an explanation of the overall scale of supersymmetry breaking.

It may be possible to provide tighter bounds on the masses by considering more specific constraints. One that has been discussed in the
literature is the bound following from the stability of deuterium \cite{agrawal, deuteron}.
The deuteron is very weakly bound and small changes in the masses will
suffice to unbind it\footnote{For example Eugene Golowich (private communication)\cite{Golowich} has estimated that if the scalar coupling is decreased by
5.2\%, the deuteron will be unbound. This is half the variation that we showed was needed to unbind the heavy elements, and would lead
to a tighter bound of 1.33 for the ratio of Eq. (11). }. This happens for more modest changes than is required for the unbinding of the rest of the elements. Since deuterium is
involved in the standard mechanisms of nucleosynthesis in the early universe and in stars, the lack of a stable deuteron could be the obstacle
to providing the elements needed for life. However, this bound is less robust that considered above. On the one hand, there may be alternate pathways
to the production of enough elements needed for life. In addition, Weinberg estimates that even an unstable
deuteron could live long enough to generate the elements \cite{weinbergmultiverse}.
Moreover, there are extra subtleties in estimating the quark-mass sensitivity of the
various two-nucleon systems \cite{Beane:2002vs, Beane2, meissner}. For all
these reasons, we consider
only the most robust of constraints, as discussed above. These strong constraints already provide very strong bounds on the masses, as summarized above.
\section*{Acknowledgement} We thank Dick Furnstahl and Brian Serot for
informative email exchanges about their results. JFD appreciated
the hospitality of IHES where most of this work was performed, and thanks
Eugene Golowich for information about weakly bound systems. This
work has been partially supported by the U.S National Science
Foundation.

\end{document}